\documentclass[12pt]{iopart}
\usepackage{iopams}
\usepackage{setstack}
\usepackage{graphicx}
\usepackage{epsfig}

\begin{document}

\title{Modified Regge calculus as an explanation of dark energy}

\author{W.M. Stuckey$^1$, T.J. McDevitt$^2$ and M. Silberstein$^3$}

\address{$^1$ Department of Physics \\ Elizabethtown College \\ Elizabethtown, PA  17022}
\address{$^2$ Department of Mathematical Sciences \\ Elizabethtown College \\ Elizabethtown, PA  17022}
\address{$^3$ Department of Philosophy \\ Elizabethtown College \\ Elizabethtown, PA  17022}

\begin{abstract}
Using Regge calculus, we construct a Regge differential equation for
the time evolution of the scale factor $a(t)$ in the Einstein-de
Sitter cosmology model (EdS). We propose two modifications to the
Regge calculus approach: 1) we allow the graphical links on spatial
hypersurfaces to be large, as in direct particle interaction when
the interacting particles reside in different galaxies, and 2) we
assume luminosity distance $D_L$ is related to graphical proper
distance $D_p$ by the equation  $D_L =
(1+z)\sqrt{\overrightarrow{D_p}\cdot \overrightarrow{D_p}}$, where
the inner product can differ from its usual trivial form. The
modified Regge calculus model (MORC), EdS and $\Lambda$CDM are
compared using the data from the Union2 Compilation, i.e., distance
moduli and redshifts for type Ia supernovae. We find that a best fit
line through $\displaystyle
\log{\left(\frac{D_L}{\mbox{Gpc}}\right)}$ versus $\log{z}$ gives a
correlation of 0.9955 and a sum of squares error (SSE) of 1.95. By
comparison, the best fit $\Lambda$CDM gives SSE = 1.79 using $H_o$ =
69.2 km/s/Mpc, $\Omega_{M}$ = 0.29 and $\Omega_{\Lambda}$ = 0.71.
The best fit EdS gives SSE = 2.68 using $H_o$ = 60.9 km/s/Mpc. The
best fit MORC gives SSE = 1.77 and $H_o$ = 73.9 km/s/Mpc using $R =
A^{-1}$ = 8.38 Gcy and $m = 1.71\times 10^{52}$ kg, where $R$ is the
current graphical proper distance between nodes, $A^{-1}$ is the
scaling factor from our non-trival inner product, and $m$ is the
nodal mass. Thus, MORC improves EdS as well as $\Lambda$CDM in
accounting for distance moduli and redshifts for type Ia supernovae
without having to invoke accelerated expansion, i.e., there is no
dark energy and the universe is always decelerating.
\end{abstract}
\noindent{\it Keywords \/}:Regge calculus, dark energy, $\Lambda$CDM, Einstein-de Sitter universe

%\maketitle

\thispagestyle{empty}

\section{Introduction}
\label{section1}

The problem of cosmological ``dark energy'' is by now well
known\cite{garfinkle}\cite{paranjape}\cite{tanimoto}\cite{clarkson}\cite{perlmutter1}\cite{bianchi}.
Essentially, redshifts and distance moduli for type Ia supernovae
indicate the universe is in a state of accelerated expansion when
analyzed using general relativistic
cosmology\cite{riess}\cite{perlmutter2}\cite{suzuki}. Specifically,
the distance moduli increase with increasing redshift faster than
predicted by general relativistic cosmology using matter alone.
Until this discovery in 1998, the so-called ``standard model of
cosmology'' was general relativistic cosmology with a perfect fluid
stress-energy tensor and an early period of inflation. Since this
leads to a decelerating expansion (except during the short, early
inflationary period), something `exotic' seemed to be required to
account for the unexpectedly large distance moduli at larger
redshifts, viz., dark energy that causes the universe to change from
deceleration to acceleration at about $z$ = 0.752 \cite{suzuki}. The
new ``standard model of cosmology,'' i.e., that with the most robust
fit to all observational data ($\Lambda$CDM), simply adds a
cosmological constant $\Lambda$ to the Einstein-de Sitter cosmology
model ($\Omega_{M}+\Omega_{\Lambda}=1$) and $\Lambda$ then provides
the mechanism for accelerated expansion, i.e., it provides the dark
energy. The ``problem'' is that our best theories of quantum physics
tell us the cosmological constant should be exactly
zero\cite{carroll} or something hideously large\cite{weinberg}, and
neither of these two cases holds in $\Lambda$CDM. Thus, one of the
most pressing problems in cosmology today is to account for the
unexpectedly large distance moduli at larger redshifts observed for
type Ia supernovae\cite{bianchi}.

The most popular attempts to explain the apparent accelerating
expansion of the universe include
quintessence\cite{weinberg}\cite{zlatev}\cite{wang} and
inhomogeneous
spacetime\cite{garfinkle}\cite{paranjape}\cite{tanimoto}\cite{clarkson}\cite{marra}
(there are even combinations of the two\cite{roos}\cite{buchert}).
Although these solutions have their critics\cite{zibin}, they are
certainly promising approaches. Another popular attempt is the
modification of general relativity (GR). These approaches, such as
f(R)
gravity\cite{bernal}\cite{nojiri}\cite{kleinert}\cite{capozziello1}\cite{capozziello2}\cite{olmo},
have stimulated much
debate\cite{flanagan}\cite{barausse}\cite{vollick}, which is a
healthy situation in science. Herein, we propose a new approach to
the modification of GR via its graphical counterpart, Regge
calculus.

Specifically, we construct a Regge differential equation for the
time evolution of the scale factor $a(t)$ in the Einstein-de Sitter
cosmology model (EdS), then we propose two modifications, both
motivated by our work on foundational
issues\cite{stuckey1}\cite{stuckey2}\cite{silberstein1}. First, we
allow spatial links of the Regge graph to be large (as defined
below) in accord with 1) our form of direct particle interaction
between sources in different galaxies and 2) the assumption that
Regge calculus is fundamental while GR is the continuous
approximation thereto. Of course, direct particle interaction in its
original form would require a modification to general relativistic
cosmology in and of
itself\cite{wheeler1}\cite{hawking}\cite{davies1}\cite{davies2}\cite{hoyle}\cite{narlikar}.
We are not concerned with saving direct particle interaction in its
original form and, indeed, one needn't accept our version thereof to
consider the modifications of GR proposed herein, i.e., empirical
motivations suffice. Second, we do not assume that luminosity
distance $D_L$ is trivially related to graphical proper distance
$D_p$ between photon receiver and emitter as it is in EdS, i.e., in
EdS $D_L = (1+z)d_p$ where $d_p$ is proper distance between photon
receiver and emitter. There are two reasons we do not make this
assumption. First, in our view, space, time and sources are
co-constructed, yet $D_p$ is found without taking into account EM
sources responsible for $D_L$. That is to say, in Regge EdS (as in
EdS) we assume that pressureless dust dominates the stress-energy
tensor and is exclusively responsible for the graphical notion of
spatial distance $D_p$. However, even though the EM contribution to
the stress-energy tensor is negligible, EM sources are being used to
measure the spatial distance $D_L$. Second, in the continuous, GR
view of photon exchange, one considers light rays (or wave fronts)
in an expanding space to find $D_L = (1+z)d_p$. In our view, there
are no ``photon paths being stretched by expanding space,'' so we
cannot simply assume $D_L = (1+z)D_p$ as in EdS. Indeed, we find the
trivial EdS relationship between luminosity distance and proper
distance holds only when $D_p$ is small on cosmological scales. In
order to generate a relationship between $D_L$ and $D_p$, we turned
to the self-consistency equation $KQ = J$ in our foundational
approach to physics\cite{stuckey2}, where $K$ is the differential
operator, $Q$ is the `field'\footnote{The interested reader is
referred to section 3 of reference \cite{stuckey2} for an
explanation of how our notion of a ``field'' is consistent with our
notion of direct particle interaction.} and $J$ is the source. Since
we want a relationship between $D_L$ and $D_p$, the `field' of
interest is a metric $h_{\alpha\beta}$ relating the graphical proper
distance $D_p$, obtained theoretically using no EM sources, to the
luminosity distance $D_L$, obtained observationally via EM sources.
The region in question (inter-nodal region between emitter and
receiver) has metric $\eta_{\alpha\beta}$ given by $ds^2 = -c^2dt^2
+ dD_p^2$, so the inner product of interest can be written
$\eta_{\alpha\beta} + h_{\alpha\beta}$ where the spatial coordinate
is $D_p$ and $h_{\alpha\beta}$ is diagonal. Since each EM source
proper is not ``stretched out'' by the expansion of space, the
spatiotemporal relationship between emitter and receiver is modeled
per this inter-nodal region alone. Thus, unlike EdS, we have no
\textit{a priori} basis in our form of direct particle interaction
to relate $D_L$ to $D_p$, so we begin with the assumption $D_L =
(1+z)\sqrt{\overrightarrow{D_p}\cdot \overrightarrow{D_p}} =
(1+z)D_p \sqrt{1 + h_{11}}$, where $\overrightarrow{D_p}=(0,D_p)$.

The specific form of $KQ =J$ that we used was borrowed from
linearized gravity in the harmonic gauge, i.e., $\partial^2
h_{\alpha\beta} = -16 \pi G (T_{\alpha\beta} - \frac{1}{2}
\eta_{\alpha\beta} T)$. We emphasize that $h_{\alpha\beta}$ here
corrects the graphical inner product $\eta_{\alpha\beta}$ in the
inter-nodal region between the worldlines of photon emitter and
receiver, where $\eta_{\alpha\beta}$ is obtained via a matter-only
stress-energy tensor. Since the EM sources are negligible in the
matter-dominated solution, we have $\partial^2 h_{\alpha\beta} = 0$
to be solved for $h_{11}$. Obviously, $h_{11} = 0$ is the solution
that gives the trivial relationship, but allowing $h_{11}$ to be a
function of $D_p$ allows for the possibility that $D_L$ and $D_p$
are not trivially related. We have $h_{11} = AD_p + B$ where $A$ and
$B$ are constants and, if the inner product is to reduce to
$\eta_{\alpha\beta}$ for small $D_p$, we have $B = 0$. Presumably,
$A$ should follow from the corresponding theory of quantum gravity,
so an experimental determination of its value provides a guide to
quantum gravity per our view of classical gravity. As we will show,
our best fit to the Union2 Compilation data gives $A^{-1}$ = 8.38
Gcy, so the correction to $\eta_{11}$ is negligible except at
cosmological distances, as expected. Essentially, we're saying the
dark energy phenomenon is an indication that $A \neq 0$ so that one
cannot simply assume the distance $D_L$ measured using EM sources
corresponds trivially to the graphical proper distance $D_p$ even
though the EM sources contribute negligibly to the stress-energy
tensor.

One might also ask about distance corrections per $h_{00}$, i.e., as
regards redshift, but since redshift distances are fractions of a
meter one wouldn't expect $h_{00}$ to be of consequence here. Of
course, there is the issue of \textit{origin} of redshift in our
approach, since typically cosmological redshift is understood to
occur \textit{between} emission and reception\cite{misner} while
clearly it must occur \textit{during} emission and reception in our
view. While we don't have photons propagating through otherwise
empty space between emitter and receiver, we do relate the reception
and emission events in null fashion through the simplices spanning
the inter-nodal region between emitter and receiver. Using the
metric in each simplex $ds^2 = -c^2dt^2 + dD_p^2$, as above, we have
$dD_p = ad\chi$, just as in EdS, although $t$ is not proper time for
the nodal observers as it is in EdS. This difference in $t$ is
accounted for in the computation of $D_p$ where it has a small
effect for the range of data in the Union2
Compilation\footnote{There is another difference between $d_p$ and
$D_p$ as computed using $\displaystyle d\chi = \frac{cdt}{a}$ that
must be considered. This will be explained in section
\ref{section2}.}. Likewise, we do not find that it leads to a
significant difference in scale factor at time of emission $a_e$ as
a function of $z$ for the data range in question. Not surprisingly,
when we compute the redshift graphically we find it is equivalent to
the special relativity (SR) result, i.e., $\displaystyle z+1 =
\sqrt{\frac{(1+V_e/c)(1+V_r/c)}{(1-V_e/c)(1-V_r/c)}}$ where $V_e$ is
the velocity of the emitter at time of emission in the
(1+1)-dimensional inter-nodal frame and $V_r$ is the velocity of the
receiver at time of reception. Using this form of redshift in the
EdS model and comparing the result to the use of cosmological
redshift in EdS, we find there is no significant difference between
the two results for distance modulus $\mu$ versus redshift $z$ well
beyond the range of the Union2 Compilation ($z < 2$, see Figure
\ref{fig0}). Therefore, we use cosmological redshift $\displaystyle
a_e = \frac{1}{1+z}$ for the computation of $D_p$, since
cosmological redshift is far simpler than the graphical alternative.

While these modifications are motivated by our work on foundational
issues, their specific mathematical instantiations are herein aimed
at explaining dark energy. Since this is our first foray into
modified Regge calculus (MORC), the specific approaches required for
explaining other GR phenomena, e.g., the perihelion shift of
Mercury, remain to be seen. A defense of MORC will not be undertaken
here, interested readers are referred to our earlier work cited
above, but a couple comments are perhaps in order. First, the
graphical lattice used herein to obtain $a(t)$ clearly violates
isotropy and is not to be understood as a literal picture of the
distribution of matter in the universe, e.g., galactic clusters,
voids, etc. In a sense, the graphical lattice we use is no coarser
an approximation than the continuum counterpart it is designed to
replace, i.e., the featureless perfect fluid model of EdS where
there is absolutely \textit{no} structure. Rather, the graphical
lattice simply provides a `mean' evolution for the scale factor
$a(t)$ in the equation for $D_p$. Second, the goal of such idealized
models is to attempt to isolate `average' geometric and/or material
features of cosmology which broadly capture kinematic properties of
the universe as a whole. Only when such models show some initial
success are explorations into departures from their simplistic
structure motivated, e.g., the inhomogeneous spacetime models cited
above. Thus, the model we present herein was designed merely to test
the possibility of replacing the continuous EdS cosmology with a
discrete, graphical counterpart based on our form of direct particle
interaction (again, for reasons unrelated to dark energy). Only upon
some success of this initial test, i.e., improving the EdS fit to
the type Ia supernova data, should we proceed to address the
commensurate questions and implications of this approach (as
outlined briefly in section \ref{section4} of this paper). We
believe the results presented herein establish precisely ``some
initial success'' and therefore justify further exploration into
this idea.

We begin in section \ref{section2} with an overview of Regge
calculus and present our temporally continuous, spatially discrete
Regge EdS equation for the time evolution of the scale factor $a(t)$
and the commensurate equation for proper distance between photon
emitter and receiver $D_p$ in a direct inter-nodal exchange. As we
will see, the spatially discrete Regge EdS equation for the time
evolution of the scale factor $a(t)$ reproduces that of EdS when
spatial links are small. Spatial links are ``small'' when the
`Newtonian' graphical velocity $v$ between spatially adjacent nodes
on the Regge graph is small compared to c, i.e., $\displaystyle
\left (\frac{v}{c} \right)^2 \ll 1$. In that case the dynamics
between adjacent spatial nodes is just Newtonian and the evolution
of $a(t)$ in Regge EdS is equal to that in EdS. Deviations in the
evolution of $a(t)$ between Regge EdS and MORC turn out to be small
(see Figure \ref{Compare}). Thus, the modification of Regge
evolution plays a relatively minor role in the MORC fits. Rather, as
we will show, the major factor in improving EdS is $D_L = (1+z)D_p
\rightarrow D_L = (1+z)\sqrt{\overrightarrow{D_p}\cdot
\overrightarrow{D_p}}$. Since Regge EdS should give EdS when used as
originally intended\cite{regge}, the proposed mechanism for EM
coupling in MORC differs from that in Regge calculus. When $v
\approx 2c$ Regge EdS encounters the ``stop point''
problem\cite{khavari}\cite{defelice}\cite{lewis}, i.e., the backward
time evolution of $a(t)$ halts, so $a(t)$ has a minimum and there is
a maximum value of $z$ for which one can find $D_p$. Of course, this
is not a real problem for Regge EdS if one is simply using it to
model EdS, since one can regularly check $v$ in the computational
algorithm and refine the size of the lattice to ensure $v$ remains
small. However, in our case the graphical approach is fundamental,
so lattice refinements are not mere mathematical adjustments, but
would constitute new `mean' configurations of matter. Of course,
such refinements are certainly required in earlier cosmological
eras, but one would expect there exists a smallest spatial scale
(associated with a smallest nodal mass) so that eventually (evolving
backwards in time) $v \approx 2c$ could not be avoided and the
minimum $a(t)$ would be reached. Thus, there are significant
deviations from our use of Regge calculus and its (originally
intended) use as a graphical approximation to GR.

In section \ref{section3} we present the fits for EdS, MORC, and
$\Lambda$CDM to the Union2 Compilation data, i.e., distance moduli
and redshifts for type Ia supernovae\cite{amanullah} (see Figures
\ref{fig8} and \ref{fig9}). We find that MORC improves EdS as much
as $\Lambda$CDM in accounting for distance moduli and redshifts for
type Ia supernovae even though the MORC universe contains no dark
energy is therefore always decelerating. While we do not need to
invoke dark energy, we do propose modifications to classical
gravity. Thus, it is a matter of debate as to which approach
($\Lambda$CDM or MORC) is better.

Of course, the success of MORC in this context does not commit one
to our foundational motives. In fact, one may certainly dismiss our
form of direct particle interaction and simply suppose that the
metric established by EM sources deviates from that of pressureless
dust at cosmological distances in a graphical approach to gravity.
Since motives are not germane to physics, we will not present
arguments for our foundational motives here. Abandoning our motives
but keeping the MORC formalism would simply result in a situation
similar to that in $\Lambda$CDM where a cosmological constant is
added to EdS for empirical reasons. That is, one could simply view
MORC as a modification of Regge calculus for empirical reasons
without buying into our story about direct particle interaction and
co-constructed space, time and sources. Motives notwithstanding, we
believe our MORC formalism may provide creative new approaches to
other long-standing problems, e.g., quantum gravity, unification,
and dark matter. We conclude in section \ref{section4} by briefly
outlining future directions and challenges for this research
program.

\begin{figure}
\begin{center}
\includegraphics[height=60mm]{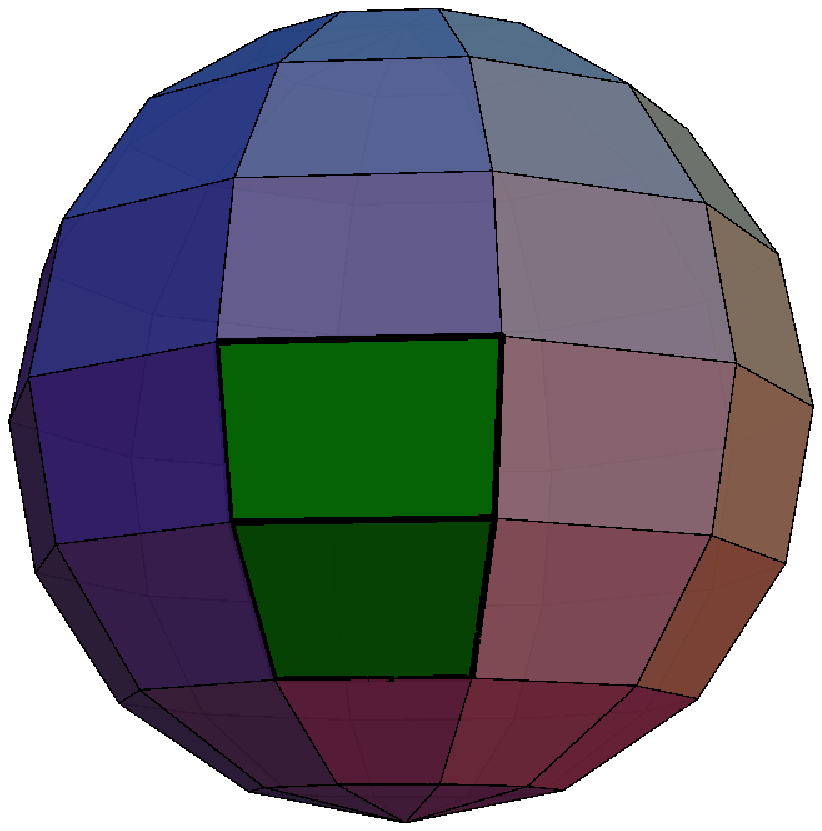} \hspace{5mm}
\includegraphics[height=60mm]{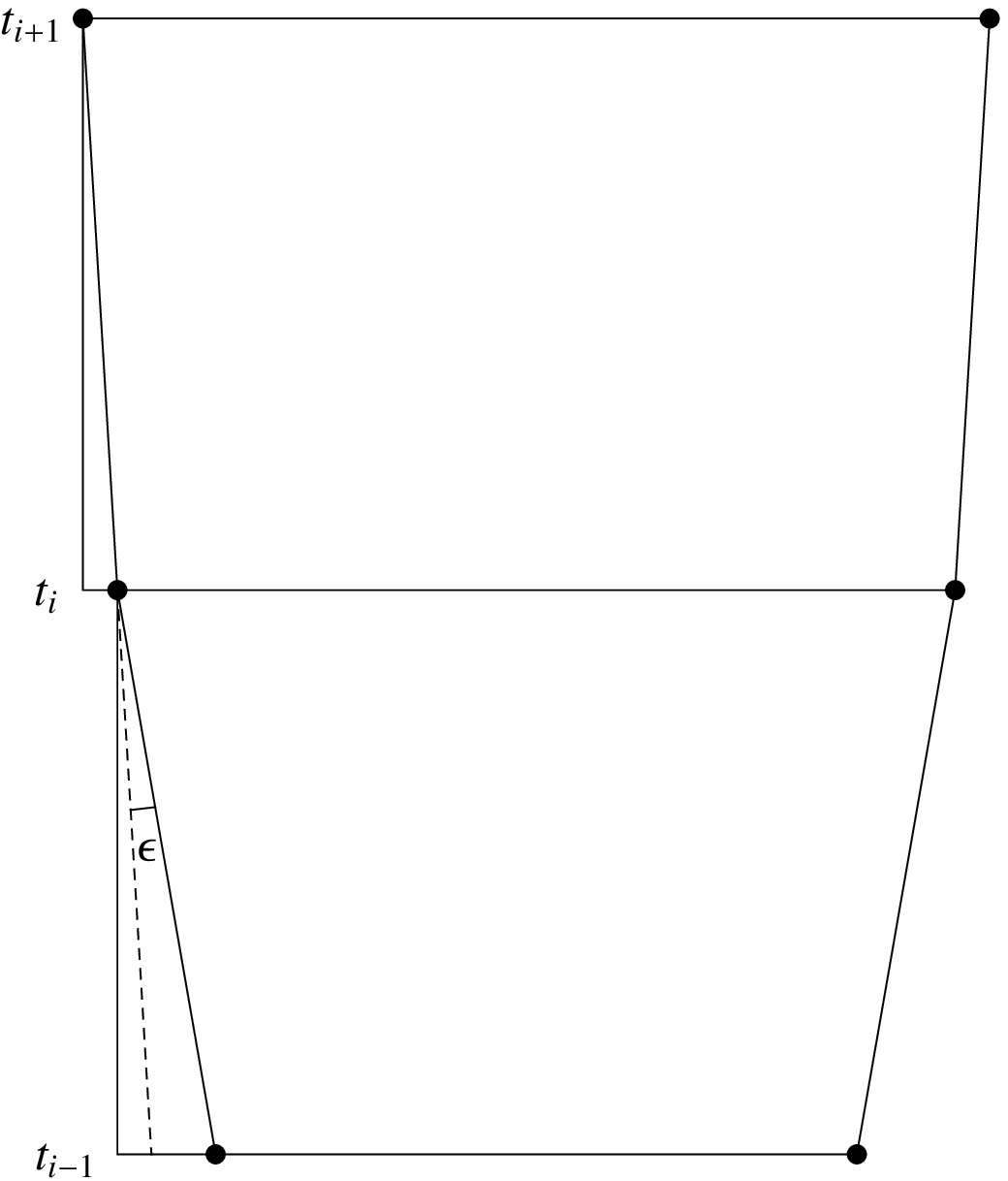}
\end{center}
\caption{(a) Tessellated sphere and  (b) two ``flattened'' trapezoids (green) from the sphere.}
\label{XY}
\end{figure}

\section{Overview of Regge Calculus}
\label{section2}

Regge calculus is typically viewed as a discrete approximation to GR
where the discrete counterpart to Einstein's equations is obtained
from the least action principal on a 4D
graph\cite{regge}\cite{misner1}\cite{barrett}\cite{williams}. This
generates a rule for constructing a discrete approximation to the
spacetime manifold of GR using small, contiguous 4D Minkowskian
graphical `tetrahedra' called ``simplices.'' The smaller the legs of
the simplices, the better one may approximate a differentiable
manifold via a lattice spacetime (Figure \ref{XY}). Although the
lattice geometry is typically viewed as an approximation to the
continuous spacetime manifold, it could be that discrete spacetime
is fundamental while ``the usual continuum theory is very likely
only an approximation\cite{feinberg}'' and that is what we assume.
Curvature in Regge calculus is represented by ``deficit angles''
(Figure \ref{XY}) about any plane orthogonal to a ``hinge''
(triangular side of a tetrahedron, which is a side of a
simplex\footnote{Our hinges are triangles, but one may use other 2D
polyhedra.}), so curvature is said to reside on the hinges. A hinge
is two dimensions less than the lattice dimension, so in 2D a hinge
is a zero-dimensional point (Figure \ref{XY}). The Hilbert action
for a vacuum lattice is $\displaystyle I_R = \frac{1}{8
\pi}\displaystyle \sum_{\sigma_i\in L}\varepsilon_{i} A_{i}$ where
$\sigma_i$ is a triangular hinge in the lattice \emph{L}, $A_i$ is
the area of $\sigma_i$ and $\varepsilon_i$ is the deficit angle
associated with $\sigma_i$. The counterpart to Einstein's equations
is then obtained by demanding $\displaystyle \frac {\delta
I_R}{\delta\ell_{j}^{2}}=0$ where $\ell_{j}^{2}$ is the squared
length of the $j^{th}$ lattice edge, i.e., the metric. To obtain
equations in the presence of matter-energy, one simply adds the
matter-energy action $I_{M}$ to $I_R$ and carries out the variation
as before to obtain $\displaystyle \frac {\delta
I_R}{\delta\ell_{j}^{2}}= -\frac {\delta I_{M}}{\delta\ell_{j}^{2}}$
\cite{sorkin}. The LHS becomes $\displaystyle \frac {\delta
I_R}{\delta\ell_{j}^{2}}=  \frac{1}{16 \pi}\displaystyle
\sum_{\sigma_i\in L}\varepsilon_{i} \cot\Theta_{ij}$ where
$\Theta_{ij}$ is the angle opposite edge $\ell_{j}$ in hinge
$\sigma_i$. One finds the stress-energy tensor is associated with
lattice edges, just as the metric, and Regge's equations are to be
satisfied for any particular choice of the two tensors on the
lattice. The extent to which Regge calculus reproduces GR has been
studied\cite{brewin1}\cite{miller1}\cite{brewin2} and general
methods for obtaining Regge equations have been
produced\cite{brewin3}, but these results are of no immediate
concern to us because we simply seek the Regge counterpart to a
specific GR equation, i.e., a Regge differential equation for the
time evolution of the scale factor $a(t)$ in EdS. Whether or not we
obtain said equation will be clear by virtue of its ability to track
the analytic EdS solution in the proper regime, so we will not have
to delve into issues associated with the `accuracy' of Regge
calculus in general.

\begin{figure}
\begin{center}
\includegraphics[height=70mm]{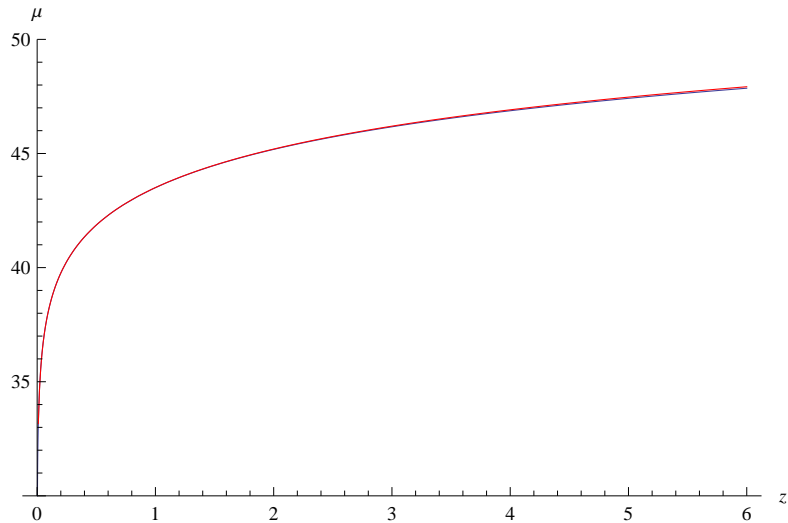}
\end{center}
\caption{Comparison of cosmological redshift (blue) and graphical special relativistic
redshift (red) using EdS. The two curves begin to be resolved at $z = 6$.}
\label{fig0}
\end{figure}

\subsection{Regge EdS Equation and MORC}

Following Brewin\cite{brewin4} and Gentle\cite{gentle1}, we take the
stress energy associated with the worldlines of our particles to be
of the form
$$
\frac {12 G m}{c^2 (ic\Delta t)}
$$
so our Regge equation is
\begin{equation*}
\fl
\frac {12 i R (a_n+a_{n+1})}{c\Delta t} \frac{ \left(\pi - \cos
^{-1} \left ( \frac {\left ( \frac Rc \right )^2 \left (\frac
{a_{n+1}-a_n}{\Delta t} \right )^2}{2 \left ( \left ( \frac Rc
\right )^2 \left (\frac {a_{n+1}-a_n}{\Delta t} \right )^2 + 2
\right ) } \right ) - 2 \cos^{-1} \left ( \frac{\sqrt{3 \left (
\frac Rc \right )^2 \left (\frac {a_{n+1}-a_n}{\Delta t} \right )^2
+ 4 }}{2 \sqrt{\left ( \frac Rc \right )^2 \left (\frac
{a_{n+1}-a_n}{\Delta t} \right )^2 + 2 }} \right ) \right
)}{\sqrt{\left ( \frac Rc \right )^2 \left (\frac
{a_{n+1}-a_n}{\Delta t} \right )^2 + 4}}
\end{equation*}
\begin{equation}
\hspace*{3.5in}
= \frac{12 i G m}{c^3 \Delta t}
   \label{Regge1}
\end{equation}
Multiplying both sides of (\ref{Regge1}) by $-ic\Delta t/12$ and
letting $v=R (a_{n+1}-a_n)/\Delta t$ gives
\begin{equation}
R (a_n+a_{n+1}) \frac{ \left(\pi - \cos ^{-1} \left ( \frac
{v^2/c^2}{2 \left ( v^2/c^2 + 2 \right ) } \right ) - 2 \cos^{-1}
\left ( \frac{\sqrt{3 v^2/c^2 + 4 }}{2 \sqrt{v^2/c^2 + 2 }} \right )
\right )}{\sqrt{v^2/c^2 + 4}} = \frac {G m}{c^2}
   \label{Regge2}
\end{equation}
If $\Delta t \rightarrow 0$, then $v$ can be regarded as a
`Newtonian' velocity and $R(a_n + a_{n+1})$ can be replaced by $2r$,
where $r$ is the graphical proper distance between two adjacent
vertices on the lattice. Equation (\ref{Regge2}) then becomes
\begin{equation}
 \frac{ \pi - \cos ^{-1} \left ( \frac {v^2/c^2}{2 \left ( v^2/c^2 + 2 \right ) } \right ) - 2 \cos^{-1} \left ( \frac{\sqrt{3 v^2/c^2 + 4 }}{2 \sqrt{v^2/c^2 + 2 }} \right )}{\sqrt{v^2/c^2 + 4}}
= \frac {G m}{2r c^2}
   \label{Regge3}
\end{equation}
which we emphasize is unmodified Regge calculus. If $v^2/c^2 \ll 1$,
then a power series expansion of the LHS of Equation (\ref{Regge3})
gives
\begin{equation}
\frac {v^2}{4c^2} + \mathcal{O} \left ( \frac vc \right )^4 = \frac
{Gm}{2rc^2}  \label{EdS}
\end{equation}
Thus, to leading order, our Regge EdS is EdS, i.e., $\displaystyle
\frac {v^2}{2} = \frac {Gm}{r}$, which is just a Newtonian
conservation of energy expression for a unit mass moving at escape
velocity $v$ at distance $r$ from mass $m$. To better understand the
relationship between Regge EdS and EdS, we note that in EdS any
comoving observer A can ask, ``What is the proper time rate of
change of proper distance for comoving observer B at a proper
distance $r$ away from me today?'' The answer is precisely $v$ given
by the EdS equation $\displaystyle \frac {v^2}{2} = \frac {Gm}{r}$,
where $m$ is the mass contained inside the sphere of radius $r$
centered on observer A. In EdS the matter is distributed uniformly
throughout space so the mass $m$ inside sphere of radius $r$ goes as
$r^3$, thus $v \propto r$ on spatial hypersurfaces in the EdS
equation, so there is no limit to how large $v$ is in this
expression, it's Newtonian. In Regge EdS, $v$ is the relative
`Newtonian' velocity of spatially adjacent nodes of mass $m$. In our
view, photon exchanges occur in direct node-to-node fashion, but
solving for a Regge graph between all galaxies in the universe is of
course unreasonable. Instead, we use Equation (\ref{Regge3}) to
provide a `mean' $a(t)$ for the computation of graphical proper
distance $D_p$ between any two photon exchangers, as in EdS, i.e.,
\begin{equation}
\mbox{proper distance} = \chi_e =  c \int_{t_e}^{t_o} \frac{dt}{a} = c \int_{a_e}^{1}\frac{da}{a\dot{a}}
 \label{propdist}
\end{equation}
We then compute $D_p$ as a function of $z$ by using Equation
(\ref{Regge3}) obtained from the `mean' graph. However, before we
continue there are two issues that we need to address regarding
Equation (\ref{propdist}).

\begin{figure}
\begin{center}
\includegraphics[height=50mm]{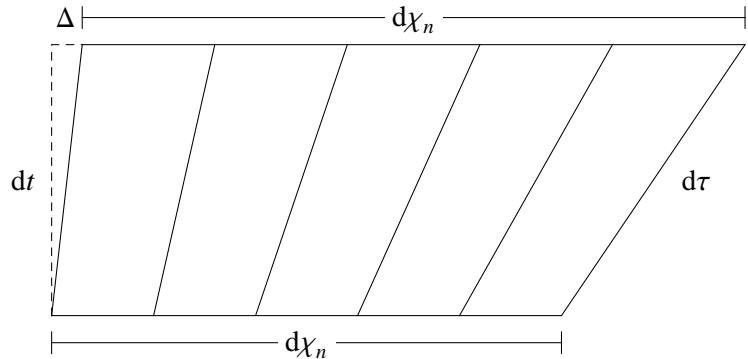}
\end{center}
\caption{Lines of constant $\chi$ are tilted away from midpoint of simplex towards emitter and receiver.}
\label{tilt}
\end{figure}

First, while it is true that $cdt = ad\chi$ for a null path in a
simplex and the null path will cross all values of $\chi$ between
emitter and receiver, the sum of $\displaystyle d\chi =
\frac{cdt}{a}$ will not equal $\chi_e$, i.e., the radial coordinate
of the emitter. That's because the lines of constant $\chi$ are
tilted in the simplices (Figure \ref{tilt}), so there is a fraction
of $d\chi$ (given by $\Delta$ in Figure \ref{tilt}) that is not
accounted for by $\displaystyle \frac{cdt}{a}$. This $\Delta$ is
positive on the emitter's side of the simplex and negative on the
receiver's side, but the $\Delta$ sum on the two sides won't cancel
out exactly, since the extent of constant-$\chi$ tilt is reduced
during the expansion. The correct equation for the graph is
\begin{equation}
\chi_e = c \int_{t_e}^{t_o}\left(1 + \frac{2V}{c}\left(\frac{\chi(t)}{\chi_e} - \frac{1}{2} \right)\right)\frac{dt}{a}
 \label{Chi}
\end{equation}
where $V$ is the SR velocity of the emitter or receiver as a
function of time and relates to our `Newtonian' $v$ per
\begin{equation}
\frac{V}{c} = \frac{v/2c}{\sqrt{1+v^2/4c^2}}
\label{Veqn}
\end{equation}
To simplify the analysis and obtain an estimate of how much $\Delta$
contributes, we use EdS with $z$ = 2 and $H_o$ = 70 km/s/Mpc. From
EdS we have $\displaystyle a(t) = \left(\frac{t}{t_o}\right)^{2/3}$,
$\displaystyle \frac{\chi}{\chi_e} = 1 -
\frac{3ct_o^{2/3}}{\chi_e}\left(t^{1/3} - t_e^{1/3}\right)$, and
$\displaystyle \frac{v}{2c} = \frac{\chi_e}{3ct_o^{2/3}t^{1/3}}$.
For $z$ = 2 and $H_o$ = 70 km/s/Mpc we have $t_o$ = 9.31 Gy, $t_e$ =
1.79 Gy, and $\chi_e$ = 11.81 Gcy. Using these values in Equation
(\ref{Chi}) we find (iteratively) $\chi_e$  = 12.189 Gcy. This
increases $\mu$ (Equation (\ref{mu}) below) by 0.069 at $z$ = 2
where $\mu$ is slightly greater than 44 (Figure \ref{fig9}). This
increase adds 0.0137 to $\displaystyle
\log{\left(\frac{D_L}{\mbox{Gpc}}\right)}$ in our curve fitting,
which amounts to a 1.3 percent increase at $z$ = 2. This change is
only 0.75 percent at $z$ = 0.5 and 0.004 percent at $z$ = 0.1. Thus,
given the scatter in the data, we will ignore this correction.

Second, in EdS, the scaling factor at emission is related to the
redshift by $\displaystyle a_e = \frac{1}{1 + z}$. In EdS, this
redshift is understood to occur while the radiation is in transit
between emitter and receiver\cite{misner}. This ``cosmological''
redshift can be understood in the graphical picture to result from
the fact that $dt$ in EdS runs along lines of constant $\chi$ and
these lines are tilted away from the center of the simplex towards
its nodal worldlines as discussed above (Figure \ref{tilt}). That
is, $\Delta = 0$ in EdS so $\displaystyle \chi_e =
\int_{t_e}^{t_o}\frac{cdt}{a}$ holds exactly. Thus, two EdS null
paths eminating from different points on a spatial link have their
proper distance of separation increase from simplex to simplex.
However, as explained above, our $dt$ is perpendicular to the
spatial links so the null paths of successive emissions do not
increase proper distance separation when traced through the
simplices, i.e., redshift occurs entirely at emission and reception.
Thus, relating successive events along the emitter's worldline in
null fashion to events on the receiver's worldline, it is not
surprising that we find the time delay between successive reception
events as related to the temporal spacing of the emission events is
that given by SR, i.e.,
\begin{equation}
z+1 = \sqrt{\frac{(1+V_e/c)(1+V_r/c)}{(1-V_e/c)(1-V_r/c)}}
\label{SRredshift}
\end{equation}
where $V_e$ is the SR velocity of the emitter at time of emission in
the (1+1)-dimensional inter-nodal frame and $V_r$ is the SR velocity
of the receiver at time of reception. Again, these SR velocities
relate to our graphical `Newtonian' $v$ per Equation (\ref{Veqn}).
As above, we simplify the analysis using the EdS equation for $a(t)$
and find $v_r = \chi_e H_o$ and $\displaystyle v_e = \frac{\chi_e
H_o}{\sqrt{a_e}}$ where, again, $\chi_e$ is the comoving coordinate
of the emitter with the receiver at the origin. We need to find
$\sqrt{a_e}$ as a function of $z$, then substitute into the equation
for proper distance between photon exchangers in EdS
\begin{equation}
d_p = \frac{2c}{H_o} \left (1 - \sqrt{a_e} \right )
\label{dpeqn}
\end{equation}
Even with the simplifications, the process gets messy and ultimately
was solved numerically. Since $a_o$ = 1, we have $d_p = \chi_e$ (as
assumed in Equation \ref{propdist}). Let $\displaystyle x =
\frac{\chi_e H_o}{2c}$ and we find
\begin{equation}
\sqrt{a_e} = x\sqrt{\frac{(A+1)^2}{(A-1)^2} - 1}
\label{aeeqn}
\end{equation}
where
\begin{equation}
A = \frac{(z+1)^2(\sqrt{1+x^2} - x)}{\sqrt{1+x^2} + x}
\label{A}
\end{equation}
Thus, Equation (\ref{dpeqn}) is $\displaystyle x = 1 -
x\sqrt{\frac{(A+1)^2}{(A-1)^2} - 1}$ and gives
\begin{equation}
A^2 - 2A + 1 - 2xA^2 + 4Ax - 2x + A^2x^2 + x^2 - 6Ax^2 = 0
\label{Acrap}
\end{equation}
We then solve Equation (\ref{Acrap}) numerically for $x$ as a
function of $z$ and compare with the EdS version, i.e.,
$\displaystyle x = 1 - \frac{1}{\sqrt{1+z}}$ to obtain Figure
\ref{fig0} where we see that there is no significant difference
between the two results well beyond the range of the Union2
Compilation ($z < 2$).

Since these two differences between MORC and EdS do not result in
any significant difference in our fit to the data of interest, we
simply use Equation (\ref{propdist}) with $\displaystyle a_e =
\frac{1}{1+z}$ to compute $D_p$. However, there is one additional
difference between $d_p$ and $D_p$ when using Equation
(\ref{propdist}) that we will not ignore. We will address this
additional (simple) correction in the following section where we fit
EdS, MORC, and $\Lambda$CDM to the Union2 Compilation.

\section{Data Analysis}
\label{section3}

The Union2 Compilation provides distance modulus $\mu$ and redshift
$z$ for each supernova. In order to find $\mu$ versus $z$ for each
model, we first find proper distance as a function of $z$, then
compute the luminosity distance $D_L$, and finally
\begin{equation}
\mu = 5 \log \left (\frac{D_L}{10 \mbox{pc}} \right ) \label{mu}
\end{equation}
For EdS we have Equation (\ref{dpeqn}) for $d_p$, so the only
parameter in fitting EdS is $H_o$. For $\Lambda$CDM we have
$\displaystyle \dot{a} = H_o\sqrt{\frac{\Omega_M}{a}+\Omega_\Lambda
a^2}$ where $\Omega_M +\Omega_\Lambda = 1$. Plugging this into
Equation (\ref{propdist}) we obtain
$$
\fl
d_p = \frac {c}{H_o \sqrt[4]{3} \sqrt[3]{\Omega_m} \sqrt[6]{\Omega_{\Lambda} }} \left [ F\left( \left . \cos ^{-1}\left(\frac{\sqrt[3]{\Omega_m} - \left(\sqrt{3}-1 \right) \sqrt[3]{\Omega_{\Lambda}}}{\sqrt[3]{\Omega_m} + \left(\sqrt{3} + 1 \right) \sqrt[3]{\Omega_{\Lambda}}} \right) \right | \frac{2+\sqrt{3}}{4} \right) - \right .
$$
\begin{equation}
   \left . F \left( \left . \cos ^{-1}\left(\frac{(z+1) \sqrt[3]{\Omega_m} -\left(\sqrt{3} - 1 \right) \sqrt[3]{\Omega_{\Lambda}}}{(z+1) \sqrt[3]{\Omega_m} + \left(\sqrt{3} + 1 \right) \sqrt[3]{\Omega_{\Lambda}}} \right) \right | \frac{2+\sqrt{3}}{4} \right ) \right ]
  \label{CDM}
\end{equation}
where $\displaystyle F(\phi|m) = \int_0^{\phi} \left ( 1-m \sin^2
\theta \right )^{-1/2} d\theta$ is the elliptic integral of the
first kind. Thus there are two fitting parameters for $\Lambda$CDM,
$H_o$ and either $\Omega_M$ or $\Omega_\Lambda$. For MORC, Equation
(\ref{Regge3}) gives us $a(\dot{a})$ rather than $\dot{a}(a)$, so we
modify Equation (\ref{propdist}) to read
\begin{equation}
D_p = R \int_{b_e}^{b_1} \frac {f'(b)}{b f(b)} \sqrt{1+\frac {b^2}4} db
\label{fb}
\end{equation}
where $b=R \dot{a}/c$,
\begin{equation}
f(b) = \frac{\sqrt{b^2+4}}{2 \left[ \pi - \cos ^{-1}\left(\frac{b^2}{2 b^2+4}\right) - 2 \cos ^{-1}\left(\frac{\sqrt{3 b^2+4}}{2 \sqrt{b^2+2}}\right)  \right ]}
\end{equation}
and $b_1$ and $b_e$ respectively solve
\begin{displaymath}
1 = \frac {Gm}{c^2 R} f(b_1) \qquad \mbox{and} \qquad a_e = \frac
{Gm}{c^2 R} f(b_e)
\end{displaymath}
The factor $\displaystyle \sqrt{1+\frac {b^2}4}$ is the correction
needed to adjust the time $dt$ in Equation (\ref{propdist}) to
proper time $d\tau$ of the nodal worldlines. [This is the ``one
additional difference between $d_p$ and $D_p$ when using Equation
(\ref{propdist})'' alluded to at the end of section \ref{section2}.]
Equation (\ref{propdist}) is then solved numerically for $D_p$ and
$D_L = (1+z)D_p\sqrt{1 + AD_p}$ as explained in section
\ref{section1}. There are three fitting parameters for MORC, the
inter-nodal coordinate $R$ on the `mean' graph, the nodal mass $m$
on the `mean' graph, and $A^{-1}$ from $h_{11}$. Specifying $m$ and
$R$ is equivalent to specifying $H_o$ in EdS, i.e., $\displaystyle
H_o = \sqrt{\frac{8\pi G\rho}{3}}$ in EdS with $\rho$ given by the
graphical values of $R$ and $m$ per $\displaystyle \frac{4}{3}\pi
R^3\rho = m$. Thus compared to EdS, MORC (as with $\Lambda$CDM) has
one additional fitting parameter $A^{-1}$, which presumably will be
accounted for ultimately by the corresponding theory of quantum
gravity.

\begin{figure}
\begin{center}
\includegraphics[height=60mm]{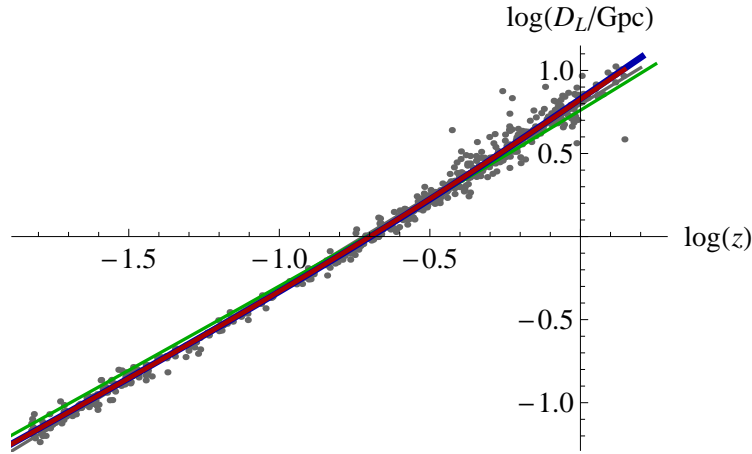}
\end{center}
\caption{Plot of transformed Union2 data along with the best fits for linear
regression (gray), EdS (green), $\Lambda$CDM (blue), and MORC (red).} \label{fig8}
\end{figure}

\begin{figure}
\begin{center}
\includegraphics[height=60mm]{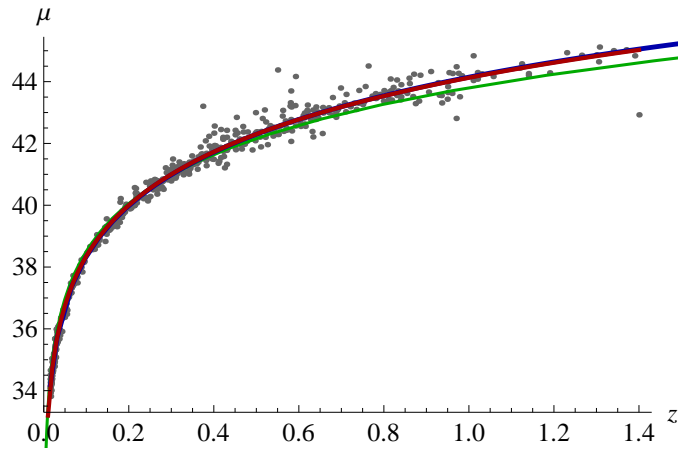}
\end{center}
\caption{Plot of Union2 data along with the best fits for EdS (green), $\Lambda$CDM (blue), and MORC (red).
The MORC curve is terminated at $z$ = 1.4 in this figure so that the $\Lambda$CDM curve is visible.}
\label{fig9}
\end{figure}

As mentioned above, we fit these three models to the Union2
Compilation data (see Figures \ref{fig8} and \ref{fig9}). In order
to establish a statistical reference, we first found that a best fit
line through $\displaystyle
\log{\left(\frac{D_L}{\mbox{Gpc}}\right)}$ versus $\log{z}$ gives a
correlation of 0.9955 and a sum of squares error (SSE) of 1.95. EdS
cannot produce a better fit than this best fit line. The best fit
EdS gives SSE = 2.68 using $H_o$ = 60.9 km/s/Mpc. A current (2011)
``best estimate'' for the Hubble constant is $H_o$ = (73.8 $\pm$
2.4) km/s/Mpc \cite{riess2}. Both MORC and $\Lambda$CDM produce
better fits than the best fit line with better values for the Hubble
constant than EdS. The best fit $\Lambda$CDM gives SSE = 1.79 using
$H_o$ = 69.2 km/s/Mpc, $\Omega_{M}$ = 0.29 and $\Omega_{\Lambda}$ =
0.71. This best fit $\Lambda$CDM is consistent with its fit to the
WMAP data using the latest distance measurements from BAO and a
recent value of the Hubble constant\cite{komatsu}. The best fit MORC
(case 1, Table 1) gives SSE = 1.77 and $H_o$ = 73.9 km/s/Mpc using
$R = A^{-1}$ = 8.38 Gcy and $m = 1.71\times 10^{52}$ kg. Given the
scatter in the data, MORC and $\Lambda$CDM produce essentially
equivalent fits, clearly superior to EdS.

\begin{figure}
\begin{center}
\includegraphics[height=50mm]{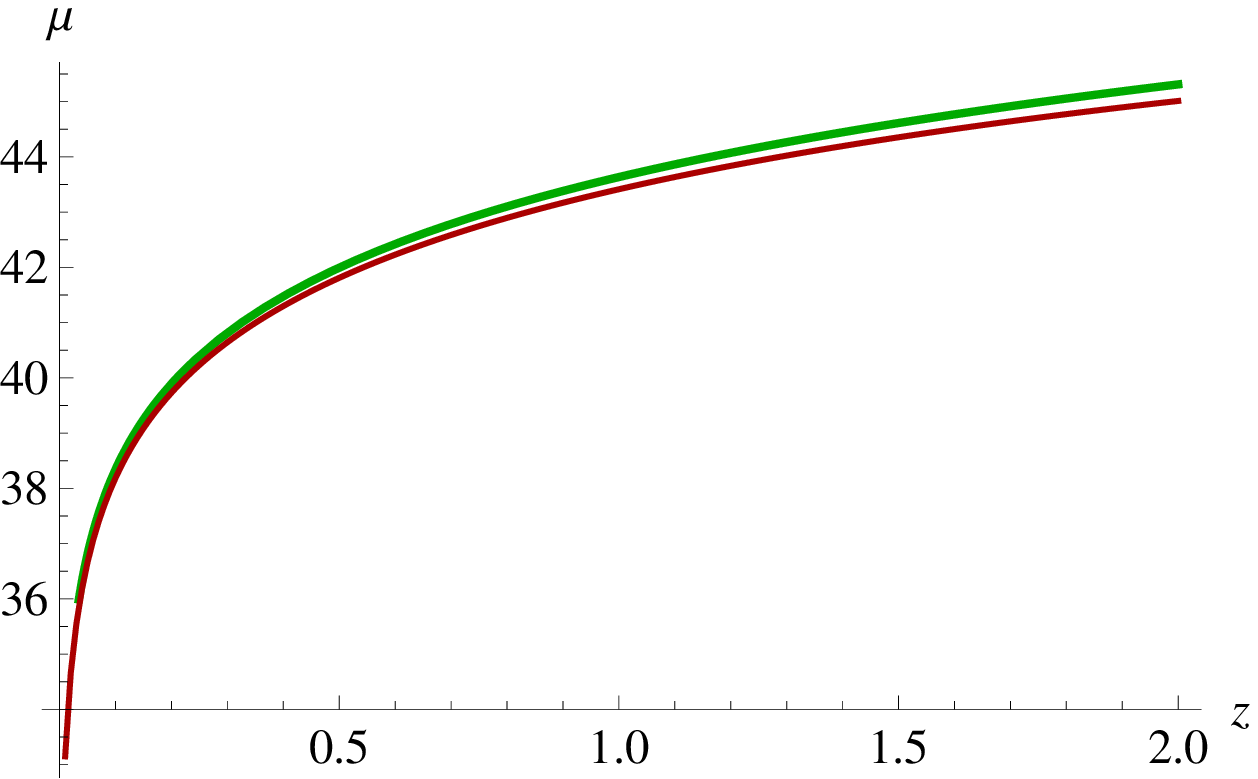} \hspace{5mm}
\includegraphics[height=50mm]{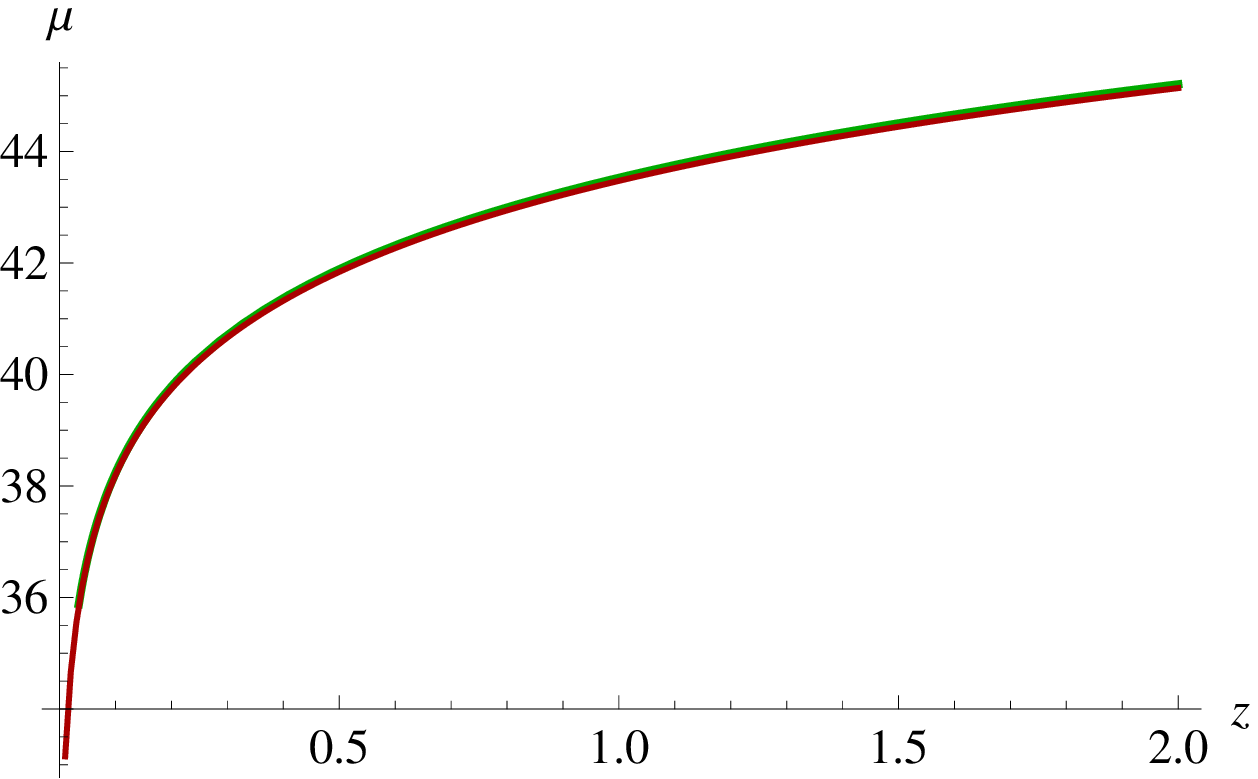}
\end{center}
\caption{Top graph shows Regge evolution (red) without $h_{11}$ correction
and EdS evolution (green) for case 1 Table 1 where the ``stop point'' $z$ is 2.05.
The bottom graph makes the same comparison for case 21 Table 1 where the ``stop point'' $z$ is 9.98.}
\label{Compare}
\end{figure}

The ``stop point'' value of $z$ in the MORC best fit is only 2.05,
so we expect the Regge evolution deviates discernibly from the EdS
evolution in this trial. To check this, we compared the Regge model
using the best fit parameters and $h_{11} = 0$ with its EdS
counterpart. As explained above, the EdS counterpart to a Regge
graphical result is obtained by using $\displaystyle H_o =
\sqrt{\frac{8\pi G\rho}{3}}$ in EdS with $\rho$ given by the
graphical values of $R$ and $m$ per $\displaystyle \frac{4}{3}\pi
R^3\rho = m$. The top graph in Figure \ref{Compare} shows there is
in fact a discernible difference between the Regge and EdS
evolutions, and the EdS value of $H_o$ obtained per $R$ and $m$ in
this trial is 68.5 km/s/Mpc, which is significantly lower than $H_o$
= 73.9 km/s/Mpc found in MORC. In fact, the twenty trials with the
lowest SSE values (cases 1-20, Table 1) have ``stop point'' $z$ less
than 10, so Regge evolution, as distinct from EdS evolution, does
come into play. However, Regge evolution tracks EdS evolution when
``stop point'' $z$ is as small 9.98 (see bottom graph in Figure
\ref{Compare}) as is true in case 21 of Table 1. And, SSE = 1.78 for
case 21 is still comparable to SSE = 1.79 of the best fit
$\Lambda$CDM. The only casuality in the higher ``stop point'' $z$
trials is $H_o$, which is lowered when Regge evolution tracks EdS
evolution. However, the $H_o$ = 71.2 km/s/Mpc in case 21 is still
comparable to $H_o$ = 69.2 km/s/Mpc for the best fit $\Lambda$CDM.
Thus, the Regge evolution plays a relatively minor role in the MORC
fits. Since we used the cosmological redshift, $\displaystyle \chi_e
= \int_{t_e}^{t_o}\frac{cdt}{a}$, and the Regge evolution played a
minor role in the MORC fits, we conclude that the major factor in
improving EdS is $D_L = (1+z)D_p \rightarrow D_L =
(1+z)\sqrt{\overrightarrow{D_p}\cdot \overrightarrow{D_p}}$. Again,
given the scatter of the Union2 Compilation data, we consider any of
the 35 MORC results in Table \ref{top50}, where SSE $\leq$ 1.78 and
$H_o$ ranges (69.9 $\rightarrow$ 75.3) km/s/Mpc, equivalent to the
best fit $\Lambda$CDM.

%\begin{figure}
%\begin{center}
%\includegraphics[height=60mm]{Figure8.eps}
%\end{center}
%\caption{Fifty lowest SS results of MORC fit.}
%\label{fig8}
%\end{figure}

\section{Discussion}
\label{section4}

We have explored a modified Regge calculus (MORC) approach to
Einstein-de Sitter cosmology (EdS), comparing the result with
$\Lambda$CDM using the Union2 Compilation of type Ia supernova data.
Our motivation for MORC comes from our approach to foundational
physics that involves a form of direct particle interaction whereby
sources, space and time are co-constructed per a self-consistency
equation. Accordingly, since EM sources are used to measure
luminosity distance $D_L$ but are not used to compute graphical
proper distance $D_p$, we did not expect $D_p$ to correspond
trivially to the luminosity distance $D_L$, i.e., we did not assume
$D_L = (1+z)D_p$. Rather, we assumed a more general relationship
$D_L = (1+z)\sqrt{\overrightarrow{D_p}\cdot \overrightarrow{D_p}}$
where the inner product employed a correction to the inter-nodal
graphical metric, $\eta_{\alpha\beta} \rightarrow \eta_{\alpha\beta}
+ h_{\alpha\beta}$ with spatial coordinate $D_p$ and
$h_{\alpha\beta}$ diagonal, so that $D_L = (1+z)D_p \sqrt{1 +
h_{11}}$. The method used to find $h_{11}$ was a form of our
self-consistency equation $KQ =J$ borrowed from the homogeneous
linearized gravity equation in the harmonic gauge, i.e., $\partial^2
h_{\alpha\beta} = 0$. While $h_{11} = 0$ is the solution typically
used, we allowed $h_{11}$ to be a function of $D_p$ which gave
$h_{11} = AD_p + B$ where $A$ and $B$ are constants. Since we wanted
the inner product to reduce to $\eta_{\alpha\beta}$ for small $D_p$,
we set $B = 0$. Our best fit MORC (case 1, Table 1) gave $A^{-1}$ =
8.38 Gcy, so the correction to $\eta_{11}$ is negligible except at
cosmological distances, as expected.

We found that in the context of the Union2 Compilation data MORC
improved EdS as well as $\Lambda$CDM without having to employ dark
energy. That is, the MORC universe evolves per pressureless dust and
is always decelerating yet it accounts for distance moduli versus
redshifts for type Ia supernovae as well as $\Lambda$CDM. Of course,
this does not commit one to our foundational motives. In fact, one
may certainly dismiss our form of direct particle interaction and
simply suppose that the metric established by EM sources deviates
from that of pressureless dust at cosmological distances; we did not
present arguments for our foundational motives here. Abandoning our
motives while keeping the MORC formalism would simply result in a
situation similar to that in $\Lambda$CDM where a cosmological
constant is added to EdS for empirical reasons, i.e., Regge calculus
was modified to account for distance moduli versus redshifts in type
Ia supernovae. Motives notwithstanding, MORC's empirical success in
dealing with dark energy gives us reason to believe this formal
approach to classical gravity may provide creative new techniques
for solving other long-standing problems, e.g., quantum gravity,
unification, and dark matter.

In order to explore this possibility, we need to check MORC against
the Schwarzschild solution, where experimental data is well
established and GR is well supported. While tests of the
Schwarzschild solution have been conducted on spatial scales much
smaller than the cosmological scales where we found a correction to
EdS, it has been shown that the simplices must be small in order to
reproduce the GR redshift and the perihelion precession of Mercury
in the Schwarzschild solution\cite{williams2}\cite{brewin5}. Thus,
we need to verify that MORC is consistent with the Schwarzschild
solution per observational data. We might refine our study of MORC
cosmology, but we feel the easiest way to test MORC is via the
Schwarzschild solution where perhaps the issue of dark matter can be
addressed in a fashion similar to dark energy in EdS. If by chance
we are able to construct a MORC for the Schwarzschild solution that
passes empirical muster, we would then consider the more general
issue of an action for modified Regge calculus in order to consider
new approaches to quantum gravity and unification. Given the level
of uncertainty involved in the next step alone, we won't speculate
further.

\begin{table}
\begin{center}
\begin{tabular}{c|ccccccc}
& $R$ & $\rho$ & $A^{-1}$ & SSE & $H_o$ & EdS $H_o$ & stop point $z$ \\
\hline
1& 25.9 & 8.15 & 25.90 & 1.77006 & 73.9081 & 65.8705 & 2.04630 \\
2& 25.9 & 8.20 & 25.90 & 1.77092 & 74.1955 & 66.0722 & 2.02772 \\
3&  25.9 & 8.10 & 25.90 & 1.77278 & 73.6205 & 65.6681 & 2.06510 \\
4&  25.9 & 8.00 & 25.95 & 1.77453 & 73.0450 & 65.2615 & 2.10341 \\
5&  25.9 & 7.95 & 25.95 & 1.77511 & 72.7569 & 65.0572 & 2.12293 \\
6&  25.9 & 8.25 & 25.90 & 1.77532 & 74.4828 & 66.2734 & 2.00937 \\
7& 25.8 & 8.45 & 25.95 & 1.77547 & 72.0349 & 67.0719 & 3.65664 \\
8& 25.8 & 8.50 & 25.95 & 1.77638 & 72.2812 & 67.2700 & 3.62925 \\
9& 25.9 & 8.35 & 25.85 & 1.77730 & 75.0570 & 66.6738 & 1.97333 \\
10& 25.8 & 8.40 & 25.95 & 1.77742 & 71.7882 & 66.8731 & 3.68436 \\
11&  25.9 & 8.05 & 25.95 & 1.77757 & 73.3328 & 65.4651 & 2.08414 \\
12& 25.7 & 8.80 & 25.95 & 1.77821 & 71.6287 & 68.4468 & 6.08675 \\
13& 25.7 & 8.75 & 25.95 & 1.77824 & 71.4054 & 68.2521 & 6.12724 \\
14& 25.9 & 8.40 & 25.85 & 1.77852 & 75.3439 & 66.8731 & 1.95563 \\
15& 25.8 & 8.65 & 25.90 & 1.77858 & 73.0178 & 67.8610 & 3.54898 \\
16& 25.9 & 8.05 & 25.90 & 1.77914 & 73.3328 & 65.4651 & 2.08414 \\
17& 25.8 & 8.70 & 25.90 & 1.77929 & 73.2626 & 68.0568 & 3.52283 \\
18& 25.9 & 7.90 & 25.95 & 1.77938 & 72.4687 & 64.8523 & 2.14270 \\
19& 25.9 & 8.30 & 25.85 & 1.77958 & 74.7700 & 66.4739 & 1.99124 \\
20& 25.8 & 8.55 & 25.95 & 1.78009 & 72.5271 & 67.4676 & 3.60218 \\
21& 25.6 & 9.00 & 25.95 & 1.78019 & 71.2375 & 69.2203 & 9.98215 \\
22& 25.6 & 8.95 & 25.95 & 1.78053 & 71.0276 & 69.0277 & 10.0435 \\
23& 25.7 & 8.85 & 25.95 & 1.78061 & 71.8515 & 68.6410 & 6.04671 \\
24& 25.8 & 8.60 & 25.90 & 1.78065 & 72.7726 & 67.6646 & 3.57542 \\
25& 25.7 & 8.70 & 25.95 & 1.78073 & 71.1816 & 68.0568 & 6.16821 \\
26& 25.5 & 9.10 & 25.95 & 1.78171 & 70.8743 & 69.6038 & 16.2143 \\
27& 25.5 & 8.90 & 26.00 & 1.78197 & 70.0626 & 68.8347 & 16.6011 \\
28& 25.6 & 9.05 & 25.95 & 1.78206 & 71.4470 & 69.4123 & 9.92147 \\
29& 25.5 & 9.15 & 25.95 & 1.78208 & 71.0759 & 69.7947 & 16.1202 \\
30& 25.6 & 8.75 & 26.00 & 1.78209 & 70.1832 & 68.2521 & 10.2959 \\
31& 25.6 & 8.80 & 26.00 & 1.78222 & 70.3950 & 68.4468 & 10.2317 \\
32& 25.4 & 9.00 & 26.00 & 1.78226 & 69.9994 & 69.2203 & 26.5859 \\
33& 25.8 & 8.35 & 25.95 & 1.78226 & 71.5412 & 66.6738 & 3.71241 \\
34& 25.3 & 9.05 & 26.00 & 1.78236 & 69.9045 & 69.4123 & 42.4792 \\
35& 25.7 & 8.55 & 26.00 & 1.78237 & 70.5076 & 67.4676 & 6.29396 \\
% 25.7 & 8.60 & 26.00 & 1.78239 & 70.7327 & 67.6646 & 6.25156 \\
% 25.0 & 9.10 & 26.00 & 1.78244 & 69.7273 & 69.6038 & 171.14300 \\
% 25.2 & 9.05 & 26.00 & 1.78245 & 69.7214 & 69.4123 & 67.90980 \\
% 25.1 & 9.10 & 26.00 & 1.78247 & 69.7999 & 69.6038 & 107.61500 \\
% 24.9 & 9.10 & 26.00 & 1.78250 & 69.6816 & 69.6038 & 271.82800 \\
% 25.4 & 8.95 & 26.00 & 1.78253 & 69.8003 & 69.0277 & 26.74000 \\
% 24.8 & 9.10 & 26.00 & 1.78257 & 69.6529 & 69.6038 & 431.40300 \\
% 24.7 & 9.10 & 26.00 & 1.78262 & 69.6347 & 69.6038 & 684.31200 \\
% 25.4 & 9.20 & 25.95 & 1.78265 & 70.7908 & 69.9852 & 25.98620 \\
% 24.6 & 9.10 & 26.00 & 1.78266 & 69.6233 & 69.6038 & 1085.15000 \\
% 24.5 & 9.10 & 26.00 & 1.78269 & 69.6161 & 69.6038 & 1720.43000 \\
% 25.3 & 9.00 & 26.00 & 1.78270 & 69.7083 & 69.2203 & 42.72070 \\
% 24.4 & 9.10 & 26.00 & 1.78270 & 69.6115 & 69.6038 & 2727.28000 \\
% 24.3 & 9.10 & 26.00 & 1.78271 & 69.6087 & 69.6038 & 4323.03000 \\
% 24.2 & 9.10 & 26.00 & 1.78272 & 69.6069 & 69.6038 & 6852.12000
\end{tabular}
\end{center}
\caption{Table of 35 trials that produced the best fits for MORC.
Column $R$ is $X$ in $R = 10^X \mbox{m}$. Column $\rho$ is $X$ in
$\rho = X\times 10^{-27} \mbox{kg/m}^3$. Column $A^{-1}$ is $X$ in
$A^{-1} = 10^X \mbox{m}$. The other columns are self explanatory.}
\label{top50}
\end{table}

\section*{References}

\end{document}